\documentstyle[aps,epsfig]{revtex}

\begin{document}
\draft

\title{Probing the $\bf \Delta NN$ Component of $\bf ^3He$}

\author{G.M.~Huber$^{(a)}$\footnotemark, G.J.~Lolos$^{(a)}$,
E.J.~Brash$^{(a)}$, S.~Dumalski$^{(a)}$, F.~Farzanpay$^{(a)}$,
M.~Iurescu$^{(a)}$, Z.~Papandreou$^{(a)}$, A.~Shinozaki$^{(a,c)}$,
A.~Weinerman$^{(a)}$, T.~Emura$^{(b)}$, H.~Hirosawa$^{(b)}$,
K.~Niwa$^{(b)}$, H.~Yamashita$^{(b)}$, K.~Maeda$^{(c)}$,
T.~Terasawa$^{(c)}$, H.~Yamazaki$^{(c)}$ S.~Endo$^{(d)}$,
K.~Miyamoto$^{(d)}$, Y.~Sumi$^{(d)}$\footnotemark, G.~Garino$^{(e)}$,
K.~Maruyama$^{(e)}$, A.~Leone$^{(f)}$, R.~Perrino$^{(f)}$,
T.~Maki$^{(g)}$, A.~Sasaki$^{(h)}$, Y.~Wada$^{(i)}$}

\address{(a) Department of Physics, University of Regina, Regina, Saskatchewan,
S4S~0A2 Canada} 
\address{(b) Department of Applied Physics, Tokyo University of
Agriculture and Technology, Koganei, Tokyo 184, Japan} 
\address{(c) Department of Physics, Tohoku University, Sendai 980, Japan} 
\address{(d) Department of Physics, Hiroshima University, Higashi-Hiroshima
724, Japan} 
\address{(e) Institute for Nuclear Study, University of Tokyo, Tanashi, Tokyo
188, Japan} 
\address{(f) INFN-Sezione di Lecce, I-73100 Lecce, Italy}
\address{(g) University of Occupational and Environmental Health, Kitakyushu
807, Japan} 
\address{(h) College of General Education, Akita University, Akita 010,
Japan} 
\address{(i) Meiji College of Pharmacy, Setagaya, Tokyo 154, Japan} 

\author{(The TAGX Collaboration)}

\date{\today}

\maketitle

\setcounter{footnote}{1}
\renewcommand{\thefootnote}{\fnsymbol{footnote}}
\footnotetext{Corresponding Author: TEL: 306-585-4240, FAX:
306-585-5659, E-mail: huberg@uregina.ca}
\setcounter{footnote}{2}
\renewcommand{\thefootnote}{\fnsymbol{footnote}}
\footnotetext{Present Address: Department of Clinical Radiology, Hiroshima
International University, Kurose-cho, Hiroshima 724-0695, Japan}

\begin{abstract}
The $^3He(\gamma,\pi^{\pm}p)$ reactions were measured simultaneously over a
tagged photon energy range of $800\le E_{\gamma}\le 1120$ MeV, well above the
$\Delta$ resonance region.  An analysis was performed to kinematically isolate
$\Delta$ knockout events from conventional $\Delta$ photoproduction events, and
a statistically significant excess of $\pi^+p$ events was identified,
consistent with $\Delta^{++}$ knockout.  Two methods were used to estimate the
$\Delta NN$ probability in the $^3He$ ground state, corresponding to the
observed knockout cross section.  The first gave a lower probability limit of
$1.5\pm 0.6\pm 0.5$\%; the second yielded an upper limit of about $2.6\%$.
\end{abstract} 


\section{Motivation}

For the last 25 years, the question of whether $\Delta$ isobars are naturally
present in the nuclear ground state with any significant probability has been
raised.  Many theoretical calculations support this conjecture.  For example, a
detailed calculation by Anastasio et al. \cite{anastasio} for the deuteron,
$^{16}O$, and infinite nuclear matter, found that the $\Delta$ content could be
as high as 7.5\%, depending on the region of Fermi momentum probed, and the
potential used to describe the $NN\rightarrow\Delta N$ transition.  Similarly,
a $NN$ and $\Delta\Delta$ coupled channels calculation with quark degrees of
freedom at short range finds it necessary to include a 5-7\% $\Delta\Delta$
content in the deuteron ground state to adequately describe the recent $T_{20}$
results from Jefferson Lab, together with the deuteron magnetic moment and $np$
scattering data \cite{lomon}.  The general consensus appears to be that it
would be surprising if nuclei had no $\Delta$ component, but the experimental
support for this remains elusive.

A fundamental limitation of all experimental searches for pre-existing
$\Delta$'s in nuclei is that assumptions must be made in the effort to link
some observed ``$\Delta$ component signal'', to the corresponding wavefunction
probability.  Each experiment is sensitive to a particular range of Fermi
momenta, while the wavefunction probability is integrated over all possible
momenta.  Therefore, the extraction of the $\Delta$ configuration probability
is model dependent, and it is not surprising that the experimental searches
have sometimes yielded contradictory results.  It is necessary to perform
studies in as many complementary manners as possible, in order to obtain a
concrete understanding of this issue.

Emura et al. \cite{chary} estimated the $\Delta NN$ component of $^3He$ as $<
2\%$, based on the asymmetry of $\pi^{\pm}p$ yields obtained with 380 to 700
MeV tagged photons.  While cuts were placed to separate conventional quasifree
$\gamma N\rightarrow\Delta$ production from any $\Delta$ knockout signal,
contamination from non-$\Delta$ processes remained in the data sample, and so
the result is not conclusive.  Amelin et al. \cite{amelin} identified
$\Delta^{++}$ knockout from $^9Be$ by 1 GeV protons, by identifying the recoil
$^8He$, and estimated the ratio of the $\Delta^{++8}He$, $p^8Li$ spectroscopic
factors to be $(6\pm 3)\times 10^{-4}$.  However, this method only probes the
quasi-bound component of the recoil system, and a further correction would have
to be attempted to yield a definitive result.  Electroproduction measurements
hold much promise, because the use of longitudinal virtual photons can provide
an effective means of separating the $\Delta$ knockout and conventional
production mechanisms, especially in a triple coincidence $(e,e'\pi^+ p)$
measurement.  A Mainz experiment \cite{blomqvist} performed a L/T separation on
the $^3He(e,e'\pi^{\pm})$ reaction at $\omega$=370 to 430 MeV, and observed an
unexpectedly large L/T ratio, which ``might be taken as a possible hint for the
existence of a preformed $\Delta$''.  Quantitative agreement with a microscopic
model, including pole terms, final state rescattering, and produced and preformed
$\Delta$ resonances, was unfortunately too poor to allow extraction of the
$\Delta$ component probability.

Perhaps the most interesting result was obtained with 500 MeV pions at LAMPF
\cite{morris,pasyuk}.  The double charge exchange reaction $(\pi^+,\pi^-p)$ can
only occur in one step if there is a pre-existing $\Delta^-$ in the nucleus, in
which case the reaction will follow quasifree
$\pi^+\Delta^-\rightarrow\Delta^0$ kinematics.  Targets ranging from $^3H$, to
$^{208}Pb$ were used, and by comparison to quasifree $(\pi^+,\pi^+p)$
scattering, $\Delta$ probabilities from 0.5 to 3.1\% were extracted.
Unfortunately, measurements were performed at only one or two angle pairs per
target, and the extracted $\Delta$ probability varied by a factor of two for
the different angle pair measurements on $^{12}C$, for example.  The method
holds much promise, but a conclusive measurement requires a systematic study
over a larger range of ejectile angles.

Here, we present the result of a recent study of the $\Delta NN$ component of
$^3He$.  This is an especially interesting nucleus for two reasons.  Firstly, a
Faddeev method calculation by Streuve et al. \cite{streuve}, explicitly
involving $NNN$ and $\Delta NN$ channels in the coupled-channel momentum-space
approach predicts the $\Delta NN$ component of the $^3He$ wavefunction to be
significant, about 2.4\%.  Secondly, any $\Delta NN$ component in $^3He$
must have unique symmetry properties, making its experimental identification
much easier.  Indeed, any $\Delta$'s present in nuclei must be deeply
off-shell, and so their existence can only be inferred on the basis of their
isospin and spin properties.  Since the $\Delta$ has $I=3/2$, the other two
nucleons are required to be in a $I=1$ state to yield $I=1/2$ for the $^3He$
ground state.  Therefore, since the two nucleons are in an isospin symmetric
state, the spin state must be the antisymmetric $^1S_0$.  This forces the
$\Delta$ to be in a $L=2$ state with respect to the $NN$ pair to give an
overall $J=1/2$ for $^3He$, and results in a unique kinematical signature,
enabling us to distinguish pre-existing $\Delta$ knockout from conventional
$\Delta$ production processes.

Furthermore, coupling the $I=3/2$ $\Delta$ with the $I=1$ $NN$ state to yield 
$I=1/2$ for $^3He$ gives the following decomposition for any $\Delta NN$ state 
in $^3He$:
\begin{displaymath}
|\Delta NN_{^3He}\rangle =\sqrt{\frac{1}{2}}|\Delta^{++}nn\rangle - 
\sqrt{\frac{1}{3}}|\Delta^+pn\rangle + \sqrt{\frac{1}{6}}|\Delta^opp\rangle .
\end{displaymath}
Thus, from isospin considerations alone, if we perform a photoproduction
experiment and identify $\pi^+ p$ from $\Delta^{++}$ decay, and $\pi^- p$ from
$\Delta^o$ decay, we anticipate a yield ratio 
\begin{displaymath}
\frac{\pi^+ p}{\pi^- p} = 9
\end{displaymath}
from $\Delta$ knockout.  In addition, the $\gamma\Delta^{++}$ interaction is
substantially stronger than the $\gamma\Delta^o$ interaction, because of the
double charge of the $\Delta^{++}$, and so this ratio is further enhanced.
These facts led Lipkin and Lee \cite{lipkin} to conclude: {\em ``Therefore, if
a strong $\pi^+$ signal and no $\pi^-$ detected in the kinematic region in
which the $\Delta$-knockout mechanism can be unambiguously identified, it is a
clear indication of the presence of $\Delta$ in $^3He$.''}

\section{Experiment}

The experiment was performed at the 1.3 GeV electron synchrotron located at the
Institute for Nuclear Study (INS) of the University of Tokyo, with tagged
photons of energy between $E_{\gamma}=800$ and 1120 MeV. This photon range is
advantageous, as the maximum of the $\gamma N\rightarrow \Delta$ quasifree
process occurs near $E_{\gamma}=350$ MeV, and so exclusive $\Delta$
photoproduction on a single nucleon is suppressed by nearly two orders of
magnitude.  The tagged photon beam was incident upon a liquid $^3He$ target,
and $\pi^+p$, $\pi^-p$ coincidences were obtained simultaneously with the TAGX
large acceptance magnetic spectrometer, with approximately $\pi$ sr solid
angle.  For more details on the TAGX system, the reader is referred to
reference \cite{nim}.  The data were obtained concurrently with
$^3He(\gamma,\pi^+\pi^-)$ results published in references
\cite{lolos,kagarlis}.  Events consisting of two charged particle
tracks, one on each side of the photon beamline, with a proton of 
momentum $> 300$ MeV/c (as reconstructed at the center of the target), in
coincidence with either a $\pi^+$ or a $\pi^-$ of momentum $> 100$ MeV/c, were
accepted for further analysis.  If more than one proton intercepted the TAGX
spectrometer, the one with the smallest scattering angle was selected, as Monte
Carlo simulations indicated that this choice was more likely to be the
$\Delta$-associated proton.  Both $\pi^+ p$ and $\pi^- p$ coincidence data were
analyzed in an identical manner, and physics-motivated criteria were placed on
the data to isolate the unique kinematical signature associated with
pre-existing $\Delta$ knockout.  These criteria are explained below.

\begin{figure}[h]
\begin{center}
\epsfig{file=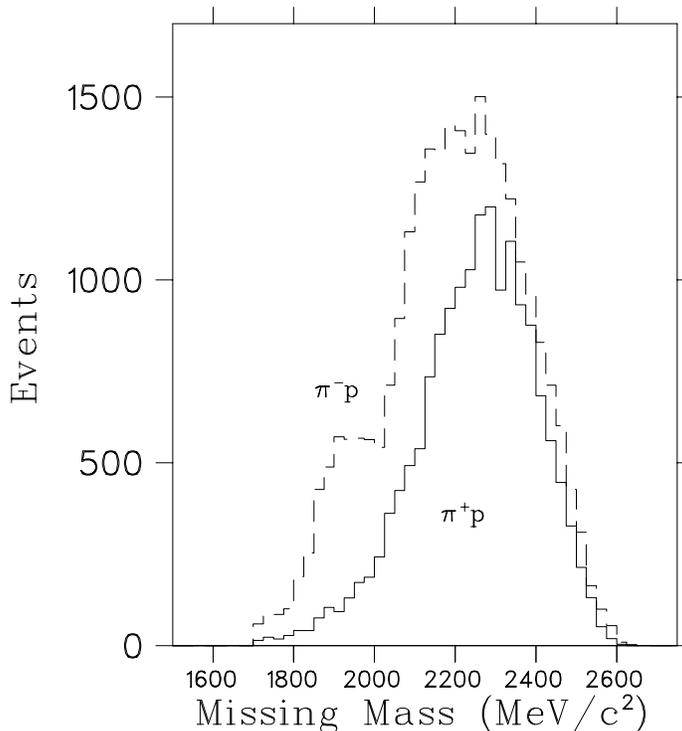,height=3.8in}~
\caption{Missing mass ($MM$) distributions for the $^3He(\gamma,\pi^{\pm}p)$
data obtained with TAGX with no ``physics cuts'' applied.  $\pi^- p$
coincidences outnumber $\pi^+ p$ over the entire histogram.  The condition $MM<
2 m_N + m_{\pi}$ was subsequently applied to eliminate events associated with
the production of a second, undetected, pion.}
\label{fig1}
\end{center}
\end{figure}

Figure 1 shows the missing mass distribution obtained with the TAGX detector.
At these energies, a significant portion of the photoabsorption cross-section
is due to $\pi\pi$ production, and events associated with the production of a
second (undetected) $\pi$ were excluded by the requirement 
\begin{equation}
Missing\ Mass (MM) < 2 m_N + m_{\pi}.
\end{equation}

It is well known that a $\pi^+p$ pair forms a pure $I=3/2$ state, while a
$\pi^- p$ pair forms a mixed $I=1/2, 3/2$ state.  The effect of this asymmetry
is shown clearly in figure 2.  Panel (a) shows the invariant mass for all of
the data, while panel (b) displays only those which passed the missing mass
requirement of equation (1).  It is observed that most of the $\pi^- p$ events
in this panel are due to the production of the various $N^*$ resonances.  As
the objective is to identify a subset of events which are due to $\Delta$
knockout, we exclude most of the remaining data with the requirement 
\begin{equation}
\pi^{\pm}p\ Invariant\ Mass (IM) \approx m_{\Delta}.
\end{equation}
These two requirements leave a small number of $\pi^{\pm}p$ events remaining, 
whose invariant mass is consistent with $\Delta$ decay, and whose missing mass
is too low to allow any undetected $\pi$ to have been produced.  

\begin{figure}[h]
\begin{center}
\epsfig{file=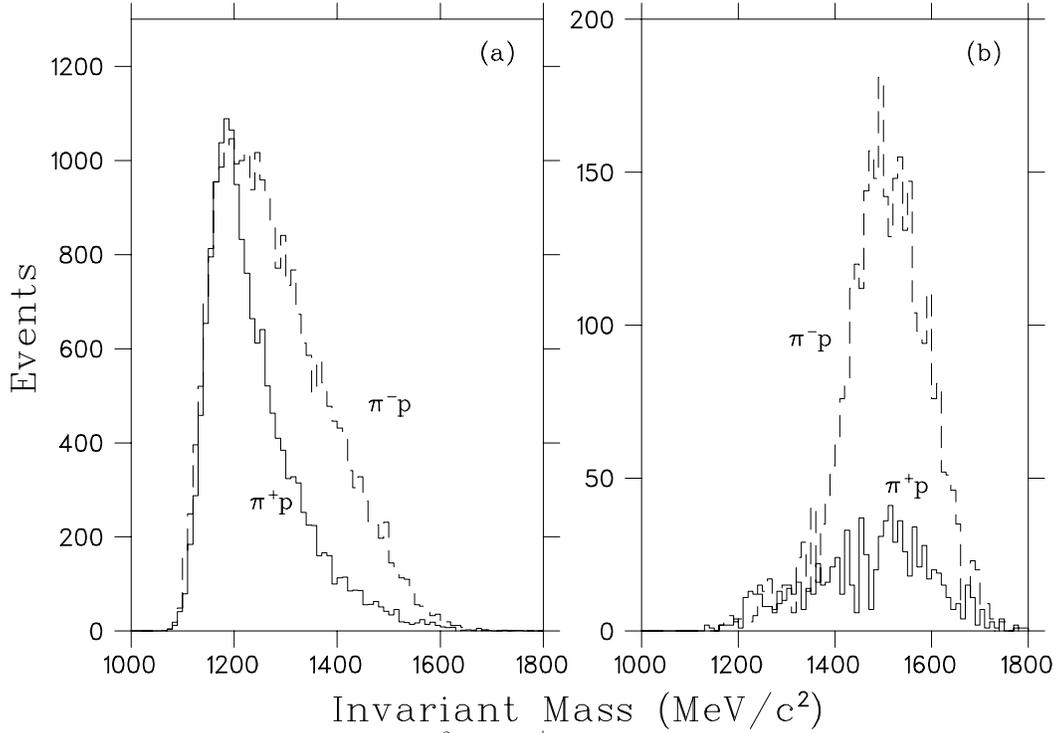,angle=90,height=3.8in}~
\caption{Invariant mass ($IM$) distributions for the $^3He(\gamma,\pi^{\pm}p)$
data obtained with TAGX.  Panel (a) has no physics cuts applied (as figure 
1), while panel (b) is subject to the requirement that $1700<MM<2020$ $\rm
MeV/c^2$.  The $\gamma n\rightarrow N^* \rightarrow p\pi^-$ channel is excluded
from the remainder of the analysis by the placement of the additional
requirement $IM\approx m_{\Delta}$.}
\label{fig2}
\end{center}
\end{figure}

The $\Delta^{++}\rightarrow \pi^+p$ channel can either be populated by
pre-existing $\Delta^{++}$ knockout, or by production processes involving more
than one nucleon, such as $\gamma pp \rightarrow \Delta^{++} n$.  The
$\Delta^0\rightarrow \pi^- p$ channel is ordinarily expected to be dominated by
quasifree $\gamma n \rightarrow \Delta^0$ production, but this is suppressed by
the choice of an incident photon energy well above the $\Delta$ region.  The
other two processes which can contribute to $\pi^- p$ production are
pre-existing $\Delta^0$ knockout, which should occur at a much lower
probability than $\Delta^{++}$ knockout, and multinucleon mechanisms such as
$\gamma np \rightarrow \Delta^0 p$.  Assuming that the isovector channel
dominates photoabsorption at these energies, we anticipate that the
multinucleon processes will contribute equally to both the $\pi^+p$ and
$\pi^-p$ channels, after accounting for the approximately 1.5 $pn$ pairs in
$^3He$. 

\begin{figure}[h]
\begin{center}
\epsfig{file=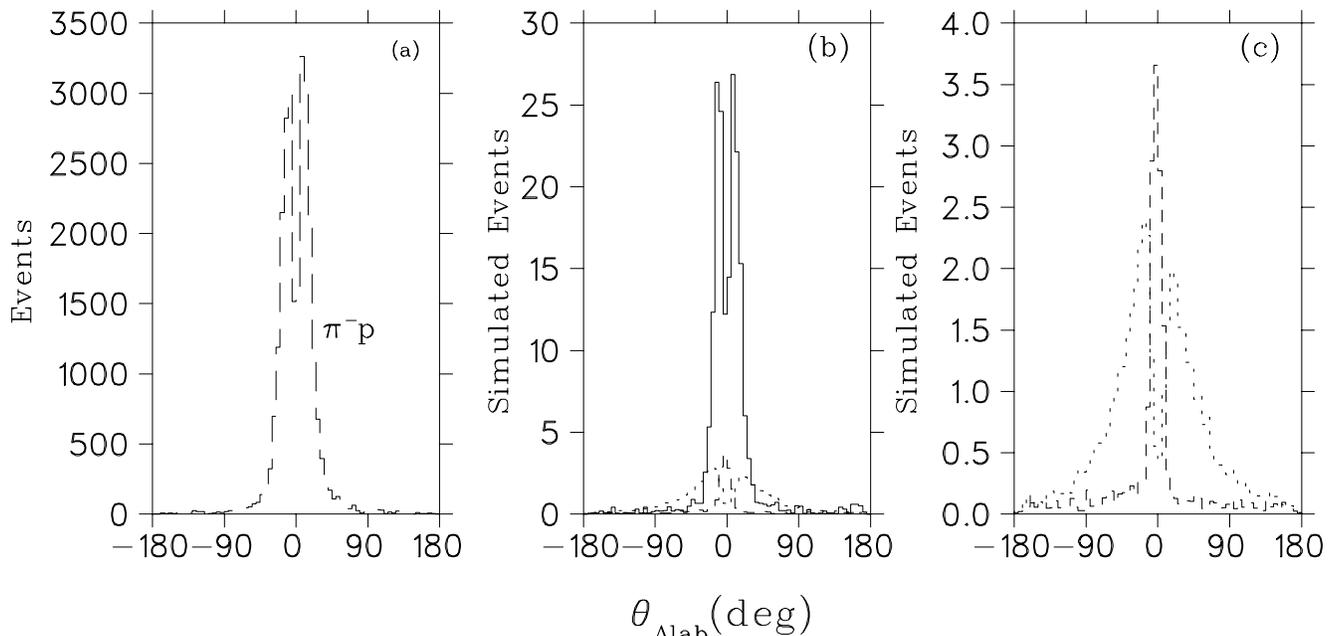,angle=90,height=3.3in}~
\caption{Reconstructed $\Delta$ emission angle in the laboratory frame for the
$\pi^-p$ channel.  Panel (a) displays the data obtained by TAGX with no physics
cuts applied.  Panel (b) shows simulations of three different production
mechanisms with arbitrary normalization.  The solid line is two pion production
such as $^3He(\gamma,\pi^o\Delta^o)pp$, and is removed from further analysis by
the missing mass cut.  Panel (c) shows closeup views of the quasifree
$\Delta^0$ photoproduction (dashed line), and $\Delta^0$ knockout (dotted line)
distributions.  The two nucleon mechanism is similar to the solid line. 
Quasifree production does not contribute to the $\pi^+p$ channel, but is
otherwise identical.  Knocked-out $\Delta$ candidate events are preferentially
selected via an additional requirement on $\theta_{\Delta lab}$.  All
simulations include the effect of the TAGX acceptance and resolution.} 
\label{fig3}
\end{center}
\end{figure}

To isolate the $\Delta$ knockout process from multinucleon $\Delta^{++}$
production, it is necessary to place an additional requirement upon the data.
Since any pre-existing $\Delta$ in $^3He$ must be in a $L=2$ state with respect
to the $NN$ pair prior to knockout, it corresponds to the high momentum
component of the wavefunction.  The angular distribution of the knocked-out
$\Delta$'s will be weighted by an additional $q^4$ factor compared to
conventionally produced $\Delta$'s from nucleons, resulting in a broad angular
distribution.  Our Monte Carlo (MC) simulations confirm that all non-knockout
mechanisms produce $\Delta$'s with forward peaked angular distributions, while
$\Delta$'s from the knockout process are distributed broadly in angle (figure
3).  This leads to the third requirement 
\begin{equation}
|\theta_{\Delta lab}|>\theta_{min},
\end{equation}
where $\theta_{min}$ is sufficiently large to discriminate against the quasifree
and two nucleon mechanisms.
\begin{figure}[h]
\begin{center}
\epsfig{file=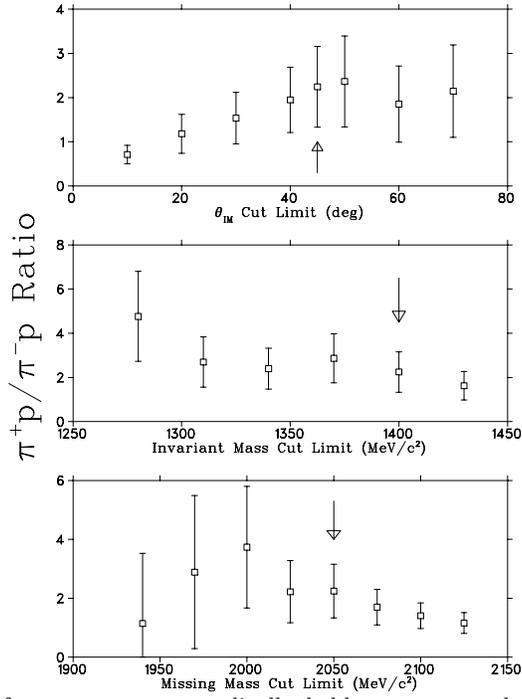,height=3.6in}
\caption{Observed $\pi^+p/\pi^-p$ ratio after two cuts are cyclically held
  constant, and the third varied.  The constantly held conditions
were $|\theta_{IM}|>45^o$, $1070<IM<1400$ $\rm MeV/c^2$, and
$1700<MM<2050$ $MeV/c^2$.  These values are denoted by the arrows on the
various plots.}
\label{fig4}
\end{center}
\end{figure}

\begin{table}
\begin{center}
\begin{tabular}{l|c|c|c}
       &     Cut Values       & $\pi^+p$ Events & $\pi^-p$ Events \\
\hline
       &$\rm 1700<MM<2025\ MeV/c^2$&                 &  \\
Narrow &$\rm 1070<IM<1370\ MeV/c^2$&  $38.2\pm 10.0$ & $9.4\pm 17.2$ \\
       &  $|\theta_{IM}|>50^o$    &                 &  \\
\hline
       &$\rm 1700<MM<2050\ MeV/c^2$&                 &  \\
Wide   &$\rm 1070<IM<1400\ MeV/c^2$&  $72.1\pm 14.5$ & $32.1\pm 22.8$ \\
       &$\rm |\theta_{IM}|>45^o$  &                 &  \\
\end{tabular}
\end{center}
\caption{The two different sets of conditions employed upon the data, and the
number of events passing each.}
\end{table}

The insensitivity of the result to the actual cuts employed is demonstrated in
figure 4, in which two cuts are cyclically held constant, and the third varied.
In the case of $\theta_{IM}$, a cut only on small angles does not adequately
remove events from the quasifree peak, but background contributions are
effectively suppressed when events with $|\theta_{IM}|>40^o$ are removed,
leading to a nearly constant ratio.  In the case of the invariant mass cut, a
narrow cut around the $\Delta$ peak results in a large excess of $\pi^+p$
events, but at the expense of poor statistics.  A broader cut results in a
diluted, but more stable ratio.  Finally, in the case of the missing mass, a
cut to exclude only the highest MM events allows in $2\pi$ background, which
dilutes the ratio.  In the end, two different sets of cuts were employed, to
allow some estimate of the systematic error in the final result.  These are
summarized in table 1.  While the data sample initially contained more $\pi^-p$
than $\pi^+ p$ events ($25411\pm 390$ versus $15733\pm 295$), the application
of conditions (1), (2) and (3) leads to a small excess of remaining $\pi^+ p$
events, compared to $\pi^- p$ events.

To see whether these $\pi^+p$ events have the kinematical signature appropriate
to $\Delta$ knockout, they were compared to Monte Carlo simulations which take
into account the TAGX resolution and acceptance and the effect of the applied
conditions (1), (2) and (3).  The modelled processes were:
\begin{enumerate}
\item{Quasifree $\gamma n\rightarrow\Delta^0$ production.  The spectral
    function of Schiavilla, Pandharipande, and Wiringa \cite{schiavilla} is
    assumed for the struck neutron, and the Breit-Wigner distribution for the
    $\Delta$.  This quasifree process does not contribute to $\pi^+p$
    production.}
\item{$\Delta$ production via two interacting nucleons, $\gamma
    NN\rightarrow\Delta N$, with the third nucleon being a spectator.  The same
    spectral function as above is assumed.}
\item{Quasifree $\gamma N\rightarrow\Delta\pi$, where the detected $\pi^{\pm}$
    may or may not originate from the $\Delta$.  Because of conditions (1) and
    (2), this reaction is expected to be excluded from the data sample, but the
    simulation was included in the analysis, to ensure that the observed events
    were not due to improperly placed cut limits.}
\item{Pre-existing $\Delta$ knockout.  Since the struck $\Delta$ can only be in
    the $L\ge 2$ state, the momentum carried by the two remaining nucleons can
    be appreciable.  The $L=2$ signature of the $\Delta$ will be smeared by the
    inelastic interaction with the incident $J=1$ photon, and so three body
    $\Delta N N$ phase space is assumed for the outgoing momentum
    distributions.  If present, it should contribute much more strongly to the
    $\pi^+p$ data set than to the $\pi^-p$ set.}
\end{enumerate}

Figure 5 shows the missing momenta of the $\pi^+p$ events passing the `narrow'
cut, and comparison with these simulations.  We see that only the
$\Delta^{++}nn$ phase space model resembles the data; the other simulated
mechanisms fail to describe the observed distribution.  While direct absorption
on three nucleons has been observed previously in photoabsorption studies on
$^3He$ \cite{emura}, a component of the photoabsorption yield with a pure
$\Delta^{++}nn$ phase space distribution has never before been reported.  Based
on this analysis, we conclude that absorption on a pre-existing $\Delta^{++}nn$
configuration in $^3He$ is the best explanation.

\begin{figure}[h]
\begin{center}
\epsfig{file=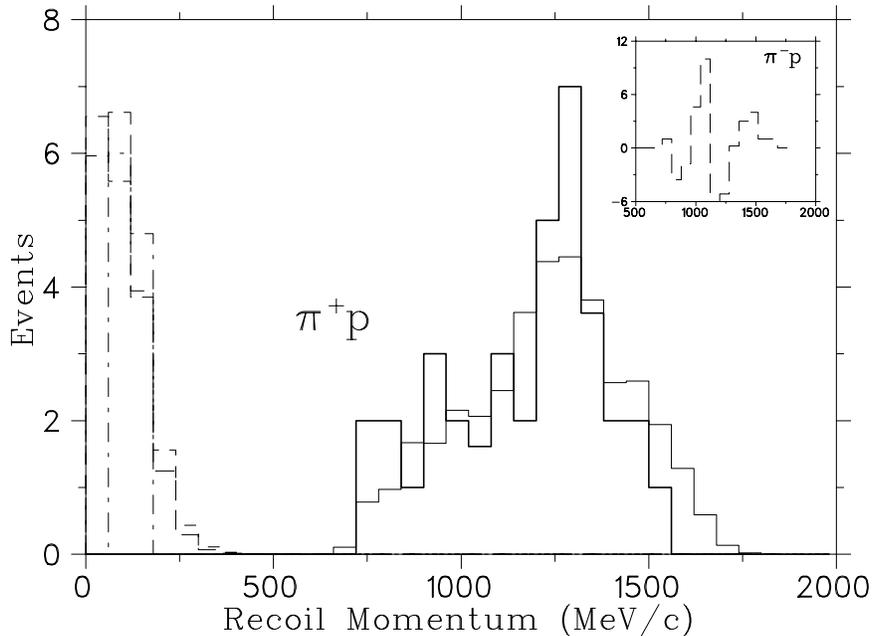,angle=90,height=3.3in}~
\caption{Missing momentum after application of `narrow' conditions (1), (2),
(3).  The dark line is the $\pi^+p$ data, while the light line is the $\Delta
NN$ phase space simulation under the same conditions.  The various dashed
distributions on the left side of the plot are the all of the other
simulations, as described in the text.  The inset shows the corresponding
$\pi^-p$ data, which is randomly distributed, and carries little statistical
significance.  The negative events are due to the empty target data
subtraction.}
\label{fig5}
\end{center}
\end{figure}

The emission angle of the $\pi^{\pm}p$ system in the frame of the
$NN$ recoil ($\theta_{\Delta NN}$) was also reconstructed for every event.
$\Delta^{++}$ originating from a $L=2$ state should be confined near $cos
\theta_{\Delta NN}=\pm 1$, while $\Delta^{++}$ due to the other processes
should be spread more uniformly in this angle.  Unfortunately, the effect of
the TAGX acceptance and conditions (1), (2), (3) is to restrict all processes
to $cos \theta_{\Delta NN} < 0.7$, but their signatures are still distinctive
enough to allow one to distinguish between them.  This is shown for the
`narrow' cut events in figure 6.  Only the $\Delta^{++}nn$ phase space
distribution is consistent in shape with the expected signature of $\Delta$
knockout, as well as with the data.

\begin{figure}[h]
\begin{center}
\epsfig{file=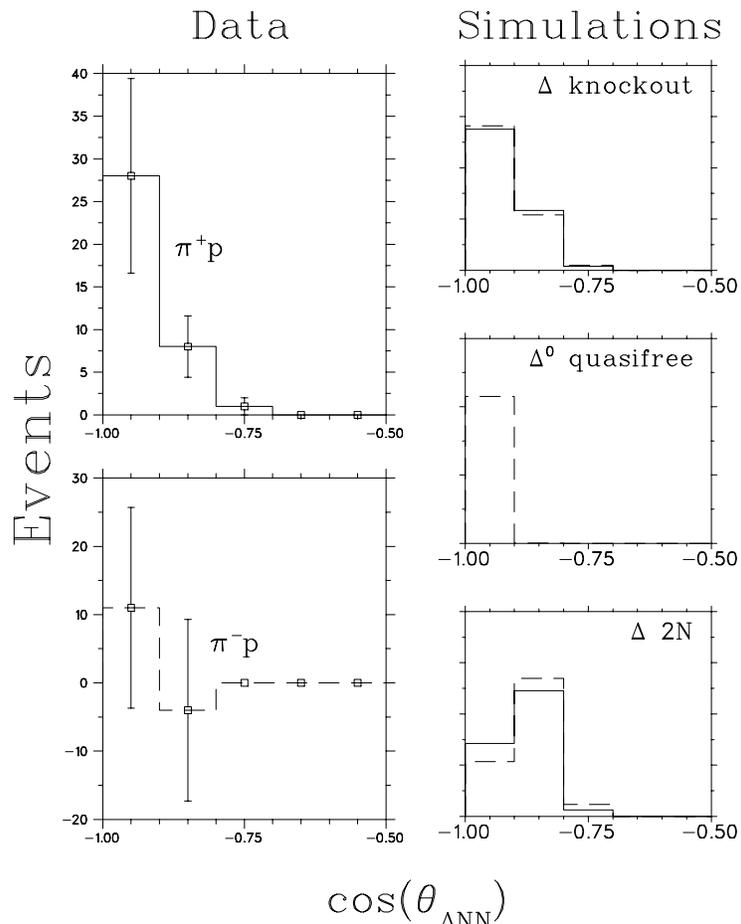,height=4.8in}~
\caption{Reconstructed $\Delta$ emission angle in the frame of the recoil $NN$
pair.  The left panels show the $^3He(\gamma,\pi^{\pm}p)$ data obtained with
TAGX after application of the cuts described in the text.  The line is to guide
the eye.  The panel on the right shows the expected distribution of events
from the various simulated production mechanisms, as labelled.  The solid
(dashed) lines indicate the $\pi^+p$ ($\pi^- p$) channels, respectively.}
\label{fig6}
\end{center}
\end{figure}

\begin{table}
\begin{center}
\begin{tabular}{l|r|r}
           & Narrow Cuts           & Wide Cuts \\
\hline
$\pi^+p$   & $0.29\pm 0.08$ $\mu$b & $0.39\pm 0.08$ $\mu$b\\
$\pi^-p$   & $0.05\pm 0.07$ $\mu$b & $0.10\pm 0.07$ $\mu$b\\
\hline
Difference & $0.24\pm 0.11$ $\mu$b & $0.29\pm 0.11$ $\mu$b \\
\end{tabular}
\end{center}
\caption{Cross sections for the events surviving the stated cuts, assuming 
phase space distribution.}
\end{table}

The phase space simulation best describes the observed $\pi^+p$ and $\pi^-p$
distributions, after application of either the wide or the narrow sets of cuts.
Therefore, to obtain the knockout cross section, the appropriately normalized 
phase space distribution was integrated over $4\pi$, leading to the cross
sections listed in table 2.  It should be noted that the acceptances for the 
two charge states are different, due to the different curvature of the charged
particle tracks in the spectrometer's magnetic field, among other factors.  If
the observed events are entirely due to $\Delta^{++}$ and $\Delta^0$ knockout,
the cross sections in table 2 should be independent of which sets of cuts are
used, and the $\pi^+p:\pi^-p$ ratio should be at least 9:1.  The results
indicate that the event sample passing the wide cuts is not entirely clean, but
that the the narrow cut sample is likely much cleaner.  A conservative estimate
of the lower limit to the $\Delta^{++}$ knockout cross section is obtained by
taking the difference between the $\pi^+p$ and $\pi^-p$ cross sections listed
in table 2.  This result is reasonably independent of which set of cuts are
used, and so the two results are averaged to obtain a $\Delta^{++}$ knockout
{\em lower limit} of $0.27\pm 0.11$ $\mu$b.

\section{Probability Estimate}

As mentioned earlier, linking the observed $\Delta^{++}$ knockout yield to the
$\Delta NN$ configuration probability requires a model.  The actual method used
varies tremendously from experiment to experiment.  Here, one possibility would
be to compare the $\Delta^{++}nn$ absorption probability to the three body
$ppn$ absorption process.  However, because the $\Delta^{++}$ is initially
highly off-shell, the inelasticities of the two processes are vastly different,
making this comparison nontrivial.  We opt to estimate the $\Delta NN$
probability in $^3He$ by comparison to the quasifree $\Delta$ photoproduction
process.  In addition to having similar inelasticities, both processes are
magnetic dipole transitions, to leading order.

If we assume the $^3He$ wavefunction to be of the form
\begin{displaymath}
|^3He\rangle = \sqrt{1-\beta^2} |ppn\rangle + \beta|\Delta NN\rangle,
\end{displaymath}
then the cross-section of the $\Delta^0$ quasifree production process is given
by 
\begin{displaymath}
|\sqrt{1-\beta^2} \langle\Delta^0pn|H_{q.f.}|ppn\rangle |^2 \frac{dN}{dE}_{q.f.}
\end{displaymath}
and the $\Delta^{++}$ knockout process by
\begin{displaymath}
|\frac{\beta}{\sqrt{2}}\langle\Delta^{++}nn|H_{k.o.}|\Delta^{++}nn\rangle |^2
\frac{dN}{dE}_{k.o.}, 
\end{displaymath}
where $dN/dE$ is the appropriate density of states factor.  In the knockout
process, the spectator $nn$ pair is in a spin antisymmetric $L=0$ state, and in
the quasifree case the $pp$ are dominantly in this same state, so it is
reasonable to expect that the spectators do not contribute to the ratio of the
two processes. 

Because quasifree $\Delta^0$ production involves a spin flip, it is dominated
by the magnetic dipole (M1) transition, and Walecka \cite{walecka} has, using
the bag model, related the transition magnetic dipole moment to the nucleon
moment 
\begin{displaymath}
\mu^*=\frac{4}{3\sqrt{2}}\mu_n=1.8\ \mu_N.
\end{displaymath}
upon substitution of the neutron magnetic moment, $\mu_n=1.91 \mu_N$ \cite{pdg}.

For $\Delta^{++}$ knockout, the E0 transition is forbidden, since the $\gamma$
has no charge, and E1 is forbidden by parity.  Thus, the leading order for
this transition should also be M1, and the ratio of the matrix elements is, to
leading order, proportional to the square of the magnetic dipole moments 
\begin{displaymath}
\frac{| \langle\Delta^{++}|H_{k.o.}|\Delta^{++}\rangle |^2}
{| \langle\Delta^0|H_{q.f.}|n\rangle |^2} = 
(\frac{\mu_{\Delta^{++}}}{\mu^*})^2.
\end{displaymath}
We will use the measurement of the $\Delta^{++}$ magnetic dipole moment from
reference \cite{bosshard}, $4.52\pm 0.50\pm 0.45$ $\mu_N$. \footnote{The
  particle data group \cite{pdg} places their estimate for $\mu_{\Delta^{++}}$
  as somewhere between 3.7 and 7.5 $\mu_N$.  However, there is a significant
  time dependence to the tabulated results, with the older experiments yielding
  higher values for the magnetic moment than the newer experiments.  We believe
  that reference \cite{bosshard} is the most reliable, as it is the newest
  measurement.}
Including both errors in quadrature leads to the ratio of the squares of the
matrix elements, above, to be $6.3\pm 1.9$.

If the quasifree $\Delta^0$ cross-section is extracted from our data sample, an
estimate of $\beta$ via the above analysis can be made.  As the incident photon
energy is well above the $\Delta$ region, some effort has to be made to isolate
the quasifree process yield from $N^*$, multipion and multinucleon production 
mechanisms.  We apply the same `narrow' condition (2) as before, and now select
forward-going $\Delta$'s via the condition 
\begin{equation}
|\theta_{\Delta lab}|<15^o,
\end{equation}
and a restrictive missing momentum cut of
\begin{equation} 
Missing\ Momentum (PM) < 185\ MeV/c 
\end{equation} 
serves to isolate the quasifree process.

Figure 7 shows the data remaining after the application of conditions (2), (4),
(5).  There is a net excess of $\pi^-p$ events in the region of opening angle
appropriate to quasifree $\Delta^0$ photoproduction.  Multipion and
multinucleon mechanisms should contribute nearly equally to the $\pi^+p$
and $\pi^-p$ yields, but the quasifree $\Delta$ photoproduction mechanism
cannot contribute to the $\pi^+p$ channel.  Assuming that all of the excess
$\pi^-p$ yield is due to quasifree $\Delta^0$ photoproduction, we obtain a
total $\Delta$ (charge integrated) photoproduction cross-section on $^3He$ of
$3.8\pm 0.4$ $\mu b$, after extrapolating over $4\pi$.  After applying a 20\%
correction for the difference in the density of states factors for the
quasifree and knockout processes (primarily due to the differing
three-momenta), and the M1 matrix elements, above, we obtain an estimate of
$\beta=0.15$, which corresponds to a lower limit on the $\Delta NN$
configuration probability of $1.5\pm 0.6\pm 0.5$ \%, where the first error
listed is statistical, and the second is due to the uncertainty in
$\mu_{\Delta^{++}}$.  A limitation of this analysis is that it does not take
into account off shell effects due to the deep binding of the $\Delta NN$
configuration.  This leads to a more compact spatial distribution, reducing the
knockout matrix element in a manner not accounted for here, and implying a
greater $\Delta NN$ configuration probability.

\begin{figure}[h]
\begin{center}
\epsfig{file=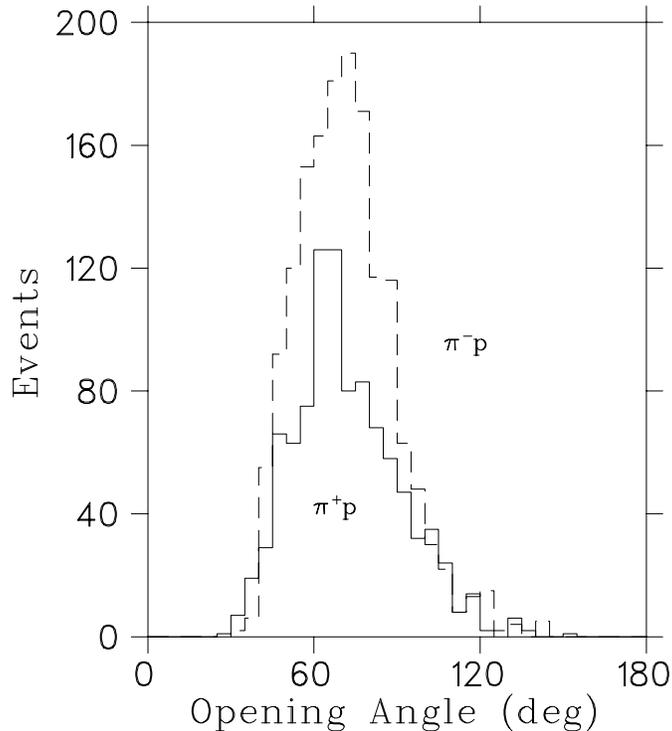,height=3.8in}~
\caption{Laboratory frame opening angle between $\pi$ and $p$ for data after
the application of conditions (2), (4), and (5), as described in the text.  An
excess of $\pi^-p$ events between opening angle of $40^o$ and $100^o$ is
observed, which corresponds to the region accessible to the quasifree process.
Outside this range, the $\pi^{\pm}p$ yields are identical.  The excess of
$\pi^-p$ over the $\pi^+p$ yield is assumed to be from quasifree $\Delta^0$
production.} 
\label{fig7}
\end{center}
\end{figure}

To check this result, the $\Delta NN$ probability is independently estimated
via an analysis of the missing momentum distribution for the $\Delta^{++}$
knockout process.  As already mentioned, the $\Delta NN$ configuration
corresponds to the high momentum component of $^3He$.  By assuming that the
high momentum tail of the spectral function of Schiavilla, Pandharipande, and
Wiringa \cite{schiavilla} is due to the $\Delta NN$ configuration, we can
obtain an upper bound on the probability of this configuration.  This spectral
function is calculated using the Faddeev method, incorporating a realistic
treatment of nucleons and deltas in nuclear matter. Reference \cite{schiavilla}
was used to form the $^3He$ Fermi momentum distribution probed by our
experiment, and normalized to unit probability.  The overlap of the $\Delta NN$
state with the $^3He$ probability distribution was obtained from the product of
the $\Delta$ knockout missing momentum distribution with the Fermi
distribution, and then normalized to the high momentum tail of the unit
probability function (figure 8).  The integral of the probability function
gives an estimated upper limit for the $\Delta NN$ probability of 2.6\%.  The
uncertainty in this method is limited by the model dependences of the assumed
momentum distributions, and is difficult to quantify.  However, the consistency
of the two results is encouraging. 

\begin{figure}[h]
\begin{center}
\epsfig{file=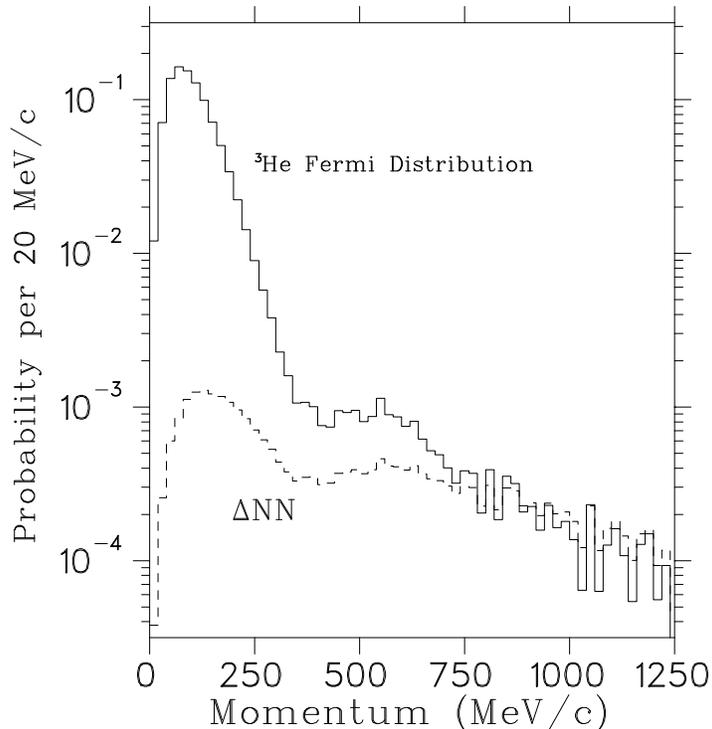,angle=90,height=3.8in}~
\caption{The solid curve is the unit probability Fermi distribution determined
from reference \protect\cite{schiavilla}.  The dashed curve is the probability
per unit momentum function for the $\Delta$ knockout process, as described in 
the text.} 
\label{fig8}
\end{center}
\end{figure}

\section{Summary}

In conclusion, we have performed an analysis to isolate $\Delta^{++}$ knockout
from $^3He$, by making use of its unique isospin and spin properties compared
to conventional photoproduction processes.  We succeeded in identifying a
kinematical region in which a small, but statistically significant number of
$\pi^+p$ events were present, and the number of $\pi^- p$ events present in the
same region was consistent with zero.  This has long been recognized as a
strong signal for pre-existing $\Delta$'s in $^3He$ \cite{lipkin}.  Based on an
electromagnetic multipole argument, and comparison to the quasifree $\Delta^0$
photoproduction process, we infer a lower limit to the $\Delta NN$ probability
in $^3He$ of $1.5\pm 0.6\pm 0.5\% $.  A second method of estimation yielded an
upper limit of 2.6\%.  We believe that the most reliable way to the extract the
$\Delta NN$ configuration probability is to measure the $\Delta$ knockout cross
section (as done here) and then use a sophisticated nuclear model to calculate
the $\Delta$ probability corresponding to this cross section.  This is the only
way that one can be assured that off-shell and Fermi momentum sampling effects
have been properly taken into account.  We encourage theorists with access to
the appropriate tools to take up this challenge.

While the result reported here has a large statistical uncertainty, isolation
of ground state $\Delta$ components is experimentally very challenging.  The
consistency of the extracted cross section for different levels of cuts, the
good agreement between data and simulations, and the fact that a $\Delta NN$
lower limit has been identified, all point to an improved measurement in this
work compared to earlier results.  Perhaps electroproduction measurements will
be able to make a more precise statement on the issue of the $\Delta$ content
of nuclei in the future.  We look forward to more stimulating results from Bonn
and Jefferson Lab over the longer term.

We wish to thank the staff of INS-ES for their considerable help with the
experiment, and Earle Lomon for his critical reading of the manuscript.  This
work was supported by grants from the Natural Sciences and Engineering Research
Council of Canada (NSERC).


\end{document}